\documentclass[runningheads]{llncs}
\usepackage[T1]{fontenc}
\usepackage{booktabs}
\usepackage{graphicx}
\usepackage{amsmath}
\usepackage{amssymb}
\usepackage{xcolor}
\usepackage{textcomp}
\usepackage{tcolorbox}
\usepackage{caption}
\usepackage{subcaption}
\usepackage{multirow}
\usepackage{floatrow}
\usepackage{float}
\floatstyle{plaintop}
\restylefloat{table}

\newcommand{\R}{\mathbb{R}}

\newcommand{\population}{50}
\newcommand{\generations}{20}
\newcommand{\maxinitlen}{eight}
\newcommand\tab[1][0.5cm]{\hspace*{#1}}

\begin{document}
\title{Search-based Optimisation of LLM Learning Shots for Story Point Estimation}
\titlerunning{Search-based Optimisation of LLM Learning Shots for SP Estimation}
%
\author{Vali Tawosi\inst{1} \and
Salwa Alamir\inst{1} \and
Xiaomo Liu\inst{2}}
\authorrunning{V. Tawosi et al.}
%
\institute{J.P.Morgan AI Research, London, UK \and
J.P.Morgan AI Research, New York, USA
\email{\{vali.tawosi,salwa.alamir,xiaomo.liu\}@jpmorgan.com}}
\maketitle              
\begin{abstract}
One of the ways Large Language Models (LLMs) are used to perform machine learning tasks is to provide them with a few examples before asking them to produce a prediction.
This is a meta-learning process known as few-shot learning. 
In this paper, we use available Search-Based methods to optimise the number and combination of examples that can improve an LLM's estimation performance, when it is used to estimate story points for new agile tasks.
Our preliminary results show that our SBSE technique improves the estimation performance of the LLM by 59.34\% on average (in terms of mean absolute error of the estimation) over three datasets against a zero-shot setting.

\end{abstract}
\section{Introduction}
Several studies proposed AI-based approaches to estimate the effort required to complete a user story in agile software development. The state-of-the-art uses deep-learning methods, leveraging the semantic similarity of the current user story to the previously estimated ones. However, the estimation performance of these methods is still inferior to baselines \cite{tawosi2022agile}, which incites further research to find more effective models. 
Recent advances with Large Language Models (LLMs) demonstrated several emergent abilities including Natural Language Understanding (NLU) at a higher level than that attained by smaller language models \cite{openai2023gpt4}.
In light of these advances, we investigated the ability of such LLMs to estimate story points for software tasks.\footnote{User stories, software tasks, and issues are used interchangeably in this paper.} 

\textbf{Problem:} During these investigations, we realised that few-shot learning, in which the LLM is provided with a few sample tasks with their estimated story points, can affect the estimation accuracy of the LLM to better or worse, depending on the samples used. On the other hand, including many samples in the prompt is not feasible (because of the prompt length limitation issues), nor is efficient (because of the cost of a long prompt, amongst other reasons). 

Therefore, we investigated the idea of using SBSE techniques to optimise the shots (i.e., the set of tasks sampled from the training set) in order to improve the LLM's estimation accuracy. To this end, we employ an uncertainty-aware multi-objective software effort estimation method called CoGEE (Confidence Guided Effort Estimation) \cite{tawosi2021multi}. We apply our proposed approach to three previously used datasets of agile software tasks, to demonstrate how effective search-based optimisation of LLM learning shots is for story point estimation. To the best of our knowledge, this is the first study using an SBSE approach to optimise learning shots for an LLM to estimate software effort.




\section{Proposed Approach}\label{sec:approach}
We adopt CoGEE, a bi-objective software development effort estimation algorithm  \cite{tawosi2021multi}, which is originally proposed to optimise regression-based effort estimation models by minimising (i) the Sum of Absolute Error (SAE) of predictions, and (ii) the Confidence Interval (CI) of the error distribution.
Instead of a regression-based model, we use CoGEE to optimise the set of examples that help an LLM achieve better estimations in a few-shot learning scenario. Thus, we define and optimise for three objectives. Two are inherited from CoGEE (i.e., SAE and CI). The third, added in this study, minimises the number of examples provided to the LLM.
It helps reduce the number of tokens per prompt, hence, reducing the cost of inference using an LLM \cite{chen2023frugalgpt}, and avoiding prompt length limitation issues
\cite{openai2023gpt4}.
Below is how we calculate each of the three fitness values.

\textbf{1) Sum of Absolute Errors (SAE)} is the sum of the distances between the actual story point values from the estimated values: \mbox{$SAE=\sum_{i=1}^{n}|a_i-e_i|$}, where $n$ is the number of user stories in the test set, $a_i$ is the actual story point value, and $e_i$ is the estimated value for the $i^{th}$ issue.

\textbf{2) Confidence Interval (CI)} is defined in Equation (\ref{eq:ci}), where the fraction is the sample standard deviation of the distribution of the absolute errors with $n$ being the size of the sample, and $\phi(p, dof)$ is the quantile function (Equation \ref{eq:qt}) which returns a threshold value $x$ below which random draws from the given cumulative distribution function would fall $p$ percent of the times \cite{hill1970algorithm}.

\begin{equation}\label{eq:ci}
    CI = \phi(p, dof) \times \frac{std(AbsoluteErrors)}{\sqrt{n-1}}
\end{equation}

\begin{equation}\label{eq:qt}
    \phi(p, dof) = inf \{x \in \R :p\leq F(x, dof)\}
\end{equation}

For a probability $0<p<1$, $F(x, dof)$ is the probability density function of a $t$-distribution function, which is a function of $x$ and $dof$ (i.e., degree of freedom) \cite{grigelionis2013student}. Confidence intervals are calculated so that this percentage is 95\%. The degree of freedom, $dof$, depends on the number of parameters we are estimating. For an $n$ sized sample, $dof=n-k$, where $k$ is the number of parameters to be estimated (here $k=1$).

\textbf{3) Number of Shots ($N$)} is the number of sample user stories from the training set provided to the LLM in a few-shot setting: $N = |E|$, where $E$ is the set of shots. Note that $E$ can be empty, leading to a zero-shot prompt.

\subsection{Computational Search}
We use a multi-objective genetic evolutionary algorithm for optimisation and implement it using {\tt pymoo} library in Python.
Below is the configuration we used.

\textbf{Problem Representation.} As we are looking for a (near)optimal subset of the training set to use as learning shots, it is natural to define the chromosome as the sequence of indexes from the training set, with each example index being a single gene. We allow for dynamic-length chromosomes with a maximum number of \maxinitlen~genes in the initial population.

\textbf{Evolutionary Operators.} A modified single-point crossover is used, which can produce zero-length offspring, as well as offspring with a length longer than \maxinitlen. Specifically, we randomly select an independent breakpoint for each of the parents $a$ and $b$ and append the first block from parent $a$ to the second block from parent $b$, and vice versa. Mutation uses three operations: (i) replace a gene with a new one (randomly sampled from the training set) with a 50\% chance, (ii) remove a randomly selected gene with a 25\% chance, and (iii) append a random new gene to the chromosome with a 25\% chance. Both cross-over and mutation operations are controlled to avoid introducing duplicate genes into the chromosome.
The mutation and crossover rates are set to 0.8 and 0.2, respectively.

\textbf{Evolutionary Algorithm.} We use NSGA-II \cite{deb2002fast}, a popular multi-objective evolutionary algorithm for optimisation. We ran the optimisation with a population of \population~individuals for \generations~generations. 
These numbers 
are lower than the usual configuration used for evolutionary algorithms. 
We justify this experimental design choice by the fact that evaluation of each individual is too expensive when it comes to multiple inferences with an LLM (note that for each individual the LLM has to estimate all the user stories in the test set). 
Nevertheless, we use small parameters to demonstrate the feasibility of the idea and call for extended research on the topic from the SBSE community.

\subsection{Estimation Model (Large Language Model)}
We use GPT-4 API from OpenAI to do the story point estimation. Specifically, we use the {\tt test-gpt4} engine, with {\tt temperature=0.0} to limit non-deterministic responses. Note that according to Ouyang et al. \cite{ouyang2023llm} this does not eliminate the risk of non-deterministic response, but minimises it. The rest of the parameters are set to their default values.

\textbf{Prompting the LLM.} A common challenge with LLMs is that the output is in the form of natural text. Therefore, to use its output in a pipeline (such as ours, where the estimation needs to be extracted from the LLM output to compute the fitness values), we need to design the prompt such that the required post-processing is minimised.
Hence, we used the prompt template provided in Fig. \ref{fig:prompt}.
The output usually is in the form of a scalar value, or a scalar value followed by `story points' (or a similar text). We used regular expressions to extract the estimated value from the LLM output. 
It is worth mentioning that we only experimented with GPT-4 as initial experiments with other models with smaller number of parameters yielded much less deterministic output. 

\begin{figure}
    \centering
    \begin{tcolorbox}[standard jigsaw, width=1\textwidth, colframe=black, before upper={\parindent15pt\noindent},opacityback=0.80,boxrule=0.5pt,arc=0pt,outer arc=0pt]
    \raggedright 
\textbf{input} $\gets$ \textit{example\_issues}[~], \textit{target\_issue}\\
\vspace{0.2cm}
\textit{prompt} = ``You are asked to estimate effort for the user story given in <>.\\
\tab \tab \tab Use \{\textit{list of SP values used in the project}\} as estimated value.''\\
\vspace{0.2cm}
\textbf{if} \textit{len}(\textit{example\_issues})>0:\\
\tab \textit{prompt} += ``A few example user stories from the same project with their estimated 
\tab \tab \tab \tab ~effort are given in the following:''\\
\tab \textbf{for each} \textit{example} \textbf{in} \textit{example\_issue}s:\\
\tab \tab \textit{prompt} += \textit{example}[\textit{text}] + ``. '' + \textit{example}[\textit{story\_points}] + `` Story Points.''\\
\vspace{0.2cm}
\textit{prompt} += ``Estimate the following user story and generate the output as a single scalar 
\tab \tab \tab ~number only, equal to the estimated story point value.''\\
\tab \tab \tab ~<'' + \textit{target\_issue}[\textit{text}] + ``>''\\  
\vspace{0.2cm}
\textbf{return} \textit{prompt}
    \end{tcolorbox}
    \caption{The prompt used with the LLM to estimate story points for a target issue, in a zero/few-shot learning (depending on the length of the example issues list).}
    \label{fig:prompt}
\end{figure}

\subsection{Dataset}
We use three projects with Jira issues published in the TAWOS dataset \cite{tawosi2022versatile}. The three projects are Appcelerator Studio (APSTUD), Apache Mesos (MESOS), and Spring XD (XD). A sample of issues from these projects is used in previous story point estimation studies \cite{tawosi2022investigating}. We use the same sample and train-test splits in this study  (refer to \cite{tawosi2022investigating} for descriptive statistics), except that our test sets include only 30 first issues from the original test sets. We use a limited number of projects and issues in this paper to minimise the cost of running LLMs.


\section{Preliminary Results}\label{sec:result}
\begin{figure}
     \centering
     \begin{subfigure}[b]{0.32\textwidth}
         \centering
         \includegraphics[width=\textwidth, trim={0 1cm 1cm 2cm}, clip]{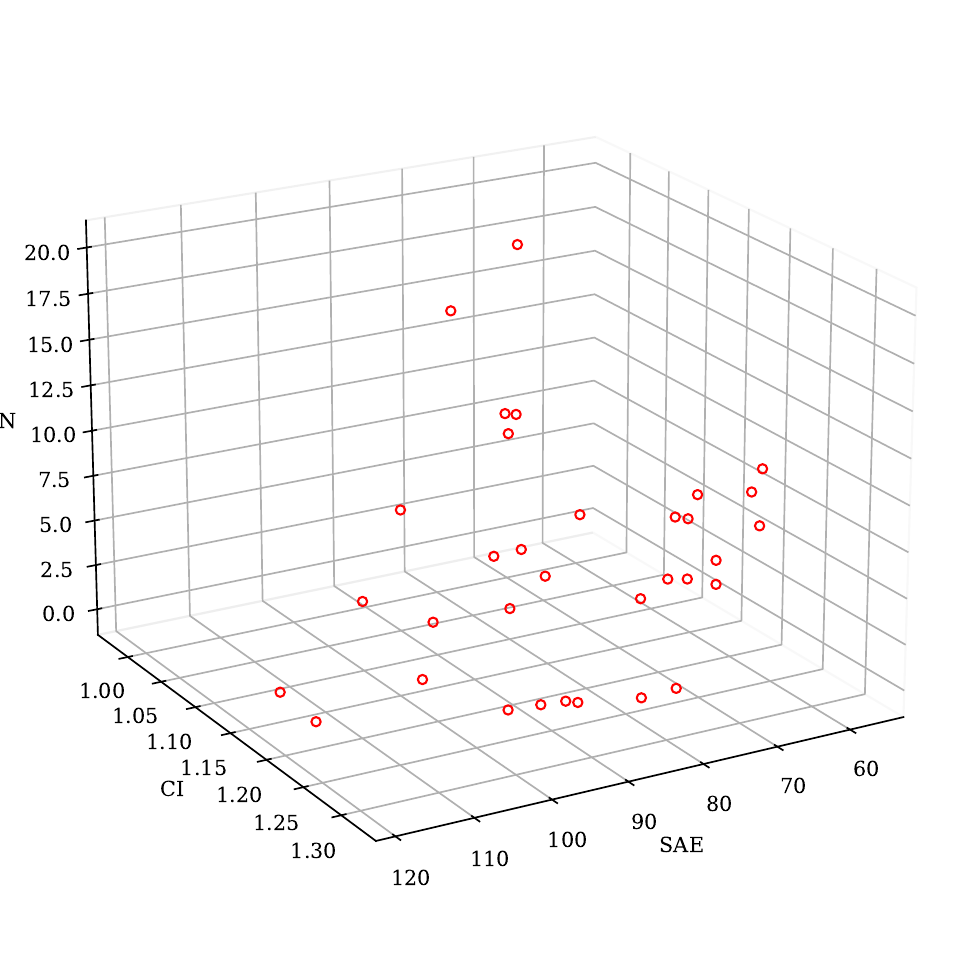}
         \caption{APSTUD}
         \label{fig:apstud pareto}
     \end{subfigure}
     \hfill
     \begin{subfigure}[b]{0.32\textwidth}
         \centering
         \includegraphics[width=\textwidth, trim={0 1cm 1cm 2cm}, clip]{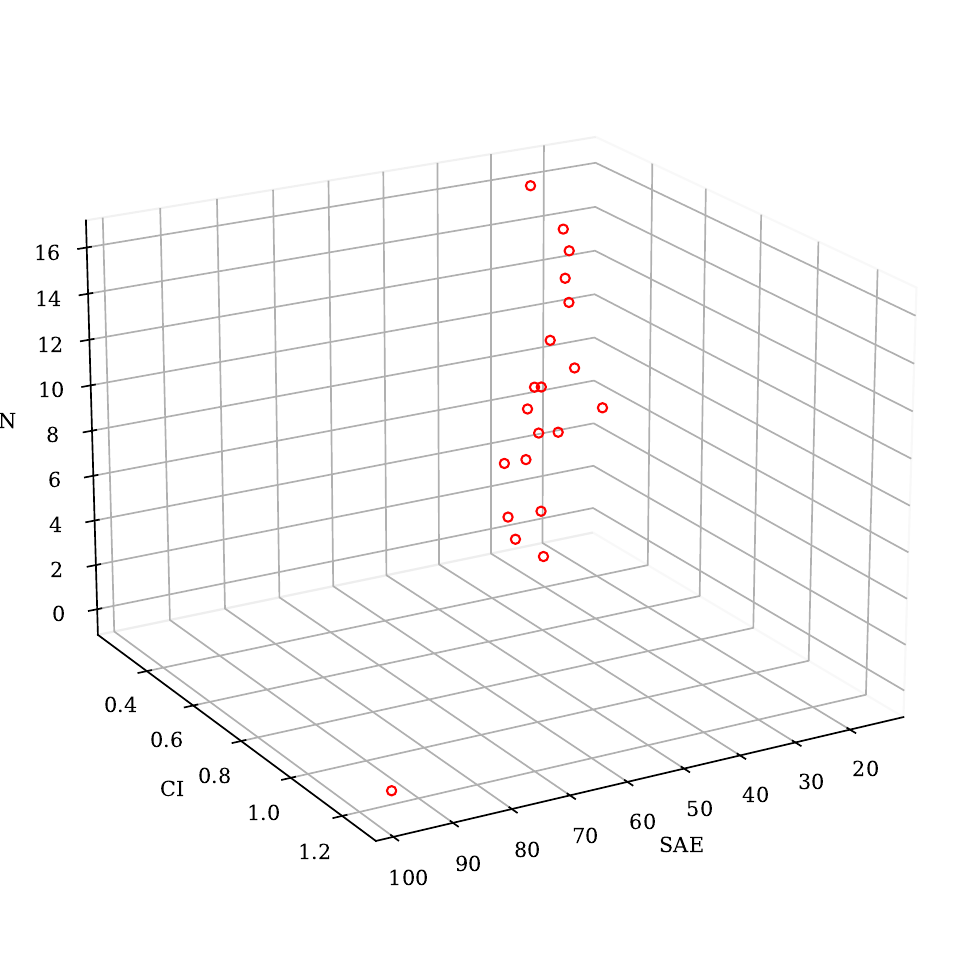}
         \caption{MESOS}
         \label{fig:messos pareto}
     \end{subfigure}
     \hfill
     \begin{subfigure}[b]{0.32\textwidth}
         \centering
         \includegraphics[width=\textwidth, trim={0 1cm 1cm 2cm}, clip]{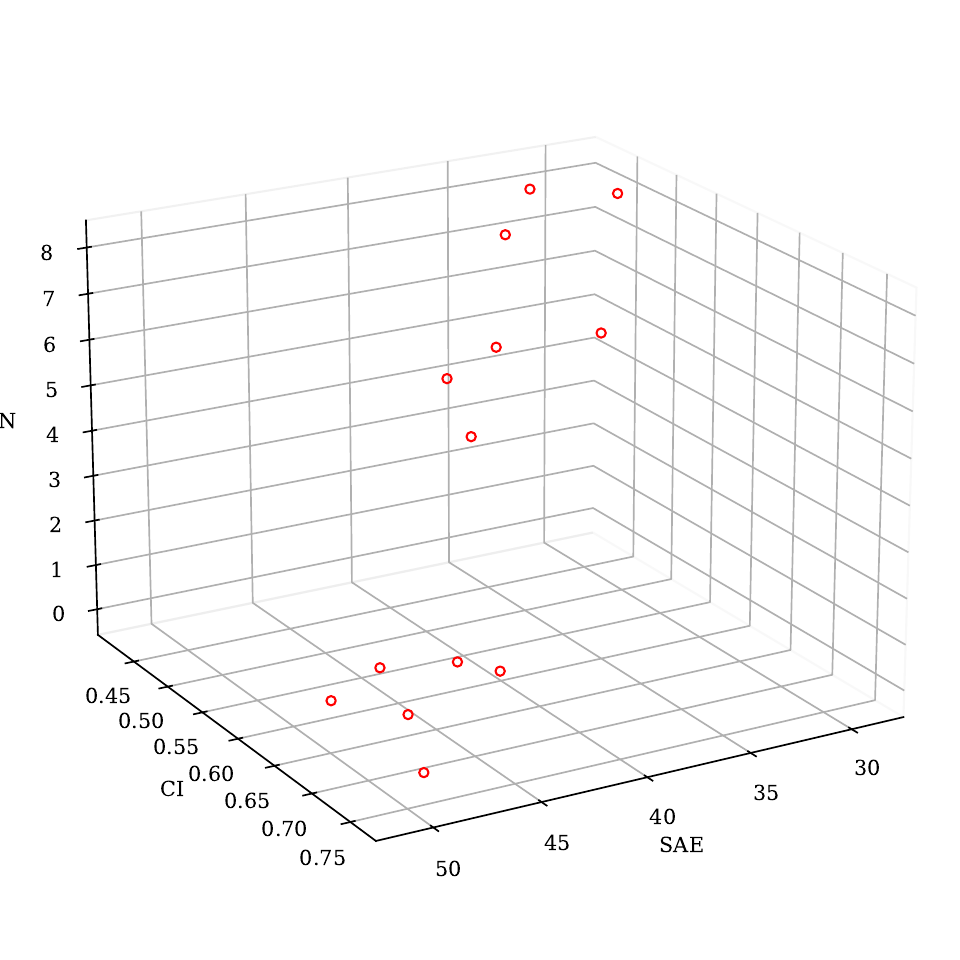}
         \caption{XD}
         \label{fig:xd pareto}
     \end{subfigure}
        \caption{Pareto fronts achieved by our proposed approach for three projects.}
        \label{fig:paretofront}
\end{figure}

Fig. \ref{fig:paretofront} shows the Pareto front formed after 20 generations of CoGEE running. It is clear that non-dominated solutions are forming a Pareto optimal front, where shots-length (N) and their combination (i.e., the specific observations selected from the train set to achieve the best performance) have a significant effect on the estimation error of the model and its level of uncertainty. In the case of MESOS, zero-shot estimation produces a large error with a wide confidence interval. However, once this individual is discarded, the column-shaped Pareto front spreads more similar to the other two projects. Overall, for all three projects, an increase in the number of shots helps the LLM to estimate better (i.e., low SAE and/or low CI). The set of non-dominated solutions provides the user with different trade-offs. One can choose to use the set of examples that minimises the number of shots to an affordable level while keeping the error and its uncertainty under an acceptable level. 

To further present the effectiveness of the proposed method, we show in Table~\ref{tab:res} the MAE (Mean Absolute Error, defined as \mbox{$\frac{SAE}{n}$}) achieved by the three sample individuals from the Pareto front, each with minimum (i.e., (near)optimal) value in one of the three objectives. 
We also provide MAE for three baseline methods including the Mean, Median, and Random Guessing baselines (see \cite{tawosi2022investigating} for definitions). 
We observe that the LLM model with zero-shot prompting
(i.e., using no samples from the previously estimated user stories) 
performs worse than Mean and Median baselines. 
In the case of MESOS, even Random Guessing performs better than the LLM with zero-shots. 
However, the individual with the lowest SAE outperforms all the other models. 

\begin{table}
\resizebox{\textwidth}{!}{
\centering
\caption{Mean absolute estimation error over the test set for sample individuals from the Pareto front achieved by CoGEE, and the baseline methods. Other objective values (i.e., N and CI) are in parenthesis. The best results are in bold.}
\label{tab:res}
\begin{tabular}{|p{2cm} | c p{2.5cm} p{3.3cm} p{3.3cm} | p{1.2cm} p{1.2cm} p{1.2cm} |}
\toprule
\multirow{2}{*}{Project} & & \multicolumn{3}{l}{Pareto Individuals with} & \multicolumn{3}{|c|}{Baselines}\\
&& Zero-shot (\textbf{N=0}) & best SAE & best CI & Mean & Median & Random \\
\midrule
APSTUD	&&	3.87~(CI=1.20)&	\textbf{1.90}~(N=9, CI=1.21)	&	2.77~(N=3, CI=\textbf{0.98})	&	2.44	&	2.27	&	5.04	\\
MESOS	&&	1.87~(CI=0.66)	&	\textbf{0.47}~(N=10, CI=0.29)	&	1.03~(N=7, CI=\textbf{0.23})	&	1.16	&	1.10&	1.77	\\
XD	&&	2.10~(CI=0.82)	&	\textbf{1.00}~(N=6, CI=0.45)	&	1.43~(N=7, CI=\textbf{0.39})	&	1.60	&	1.60	&	2.52	\\
\bottomrule
\end{tabular}}
\end{table}

\section{Conclusion}\label{sec:conclusion}

In this paper, we demonstrated promising results using SBSE techniques to improve the effectiveness of the GPT-4 model for story point estimation, via optimising the set of examples the LLM should be provided with in a few-shot learning setting.
The same can be applied to tune LLMs for any downstream task that leverages LLMs in a few-shot setting. 
In this paper, we used story point estimation as an example of a common software engineering task for demonstration purposes. The same idea can be applied to any downstream task that leverages LLMs in a few-shot setting. 
Future work can also consider minimisation of the number of tokens used per prompt, which will not only optimise to use fewer user stories as examples but also will prefer shorter user stories. This will help save even more tokens in the prompt.

Although we experimented with a limited number of projects and issues in this study, the preliminary result attests to the feasibility of the idea and invites future work to extend the study using a larger dataset and more LLMs.\\

\section*{Disclaimer} 
This paper was prepared for informational purposes by the Artificial Intelligence Research group of JPMorgan Chase \& Co and its affiliates (“JP Morgan”), and is not a product of the Research Department of JP Morgan. JP Morgan makes no representation and warranty whatsoever and disclaims all liability, for the completeness, accuracy or reliability of the information contained herein. This document is not intended as investment research or investment advice, or a recommendation, offer or solicitation for the purchase or sale of any security, financial instrument, financial product or service, or to be used in any way for evaluating the merits of participating in any transaction, and shall not constitute a solicitation under any jurisdiction or to any person, if such solicitation under such jurisdiction or to such person would be unlawful.


%
%
%
\bibliographystyle{splncs04}
\bibliography{references}

\begin{thebibliography}{10}
\providecommand{\url}[1]{\texttt{#1}}
\providecommand{\urlprefix}{URL }
\providecommand{\doi}[1]{https://doi.org/#1}

\bibitem{chen2023frugalgpt}
Chen, L., Zaharia, M., Zou, J.: {FrugalGPT}: How to use large language models
  while reducing cost and improving performance. arXiv preprint
  arXiv:2305.05176  (2023)

\bibitem{deb2002fast}
Deb, K., Pratap, A., Agarwal, S., Meyarivan, T.: A fast and elitist
  multiobjective genetic algorithm: {NSGA-II}. IEEE Transactions on
  Evolutionary Computation  \textbf{6}(2),  182--197 (2002)

\bibitem{grigelionis2013student}
Grigelionis, B.: Student's t-distribution and related stochastic processes.
  Springer (2013)

\bibitem{hill1970algorithm}
Hill, G.W.: Algorithm 396: Student's t-quantiles. Communications of the ACM
  \textbf{13}(10),  619--620 (1970)

\bibitem{openai2023gpt4}
OpenAI: {GPT-4} technical report (2023)

\bibitem{ouyang2023llm}
Ouyang, S., Zhang, J.M., Harman, M., Wang, M.: {LLM} is like a box of
  chocolates: the non-determinism of {ChatGPT} in code generation. arXiv
  preprint arXiv:2308.02828  (2023)

\bibitem{tawosi2022versatile}
Tawosi, V., Al-Subaihin, A., Moussa, R., Sarro, F.: A versatile dataset of
  agile open source software projects. In: Proceedings of the 19th
  International Conference on Mining Software Repositories. pp. 707--711 (2022)

\bibitem{tawosi2022investigating}
Tawosi, V., Al-Subaihin, A., Sarro, F.: Investigating the effectiveness of
  clustering for story point estimation. In: 2022 IEEE International Conference
  on Software Analysis, Evolution and Reengineering (SANER). pp. 827--838. IEEE
  (2022)

\bibitem{tawosi2022agile}
Tawosi, V., Moussa, R., Sarro, F.: Agile effort estimation: Have we solved the
  problem yet? insights from a replication study. IEEE Transactions on Software
  Engineering  \textbf{49}(4),  2677--2697 (2022)

\bibitem{tawosi2021multi}
Tawosi, V., Sarro, F., Petrozziello, A., Harman, M.: Multi-objective software
  effort estimation: A replication study. IEEE Transactions on Software
  Engineering  \textbf{48}(8),  3185--3205 (2021)

\end{thebibliography}
\end{document}